\def\IDvalue{SU}
\def\DIRvalue{Summary}
\def\titlevalue{Localization techniques in quantum field theories}
\def\authorvalue{Vasily Pestun$^1$, Maxim Zabzine$^2$, Francesco Benini$^{3,4}$, Tudor Dimofte$^5$,\\ Thomas T. Dumitrescu$^6$, 
Kazuo Hosomichi$^7$, Seok Kim$^8$, Kimyeong Lee$^9$, \\
Bruno Le Floch$^{10}$, 
Marcos Mari\~{n}o$^{11}$, Joseph A. Minahan$^2$, David R. Morrison$^{12}$, \\
Sara Pasquetti$^{13}$, Jian Qiu$^{2,14,15}$, Leonardo Rastelli$^{16}$, Shlomo S. Razamat$^{17}$, \\
 Silvu S. Pufu$^{18}$, Yuji Tachikawa$^{19}$, 
 Brian Willett$^{20}$, Konstantin Zarembo$^{21,2}$}

\def\addressvalue{$^1$Institut des Hautes \'Etudes Scientifique, France\\
$^2$Department of Physics and Astronomy, Uppsala University, Sweden\\
$^{3}$International School for Advanced Studies (SISSA),
via Bonomea 265, 34136 Trieste, Italy \\
$^{4}$Blackett Laboratory, Imperial College London,
London SW7 2AZ, United Kingdom \\
$^5$Perimeter Institute for Theoretical Physics, 31 Caroline St. N, Waterloo, ON N2J 2Y5, Canada\\
(on leave) Department of Mathematics, University of California, Davis, CA 95616, USA\\
$^6$Department of Physics, Harvard University, Cambridge, MA 02138, USA\\
$^7$Department of Physics, National Taiwan University,
  Taipei 10617, Taiwan\\
  $^8$Department of Physics and Astronomy \& Center for
Theoretical Physics,\\
Seoul National University, Seoul 151-747, Korea\\
$^9$School of Physics, Korea Institute for Advanced Study,
Seoul 130-722, Korea\\
$^{10}$Princeton Center for Theoretical Science, Princeton University,
Princeton, NJ 08544, USA \\
$^{11}$D\'epartement de Physique Th\'eorique et Section de Math\'ematiques\\
Universit\'e de Gen\`eve, Gen\`eve, CH-1211 Switzerland\\
$^{12}$Departments of Mathematics and Physics, 
University of California, Santa Barbara\\
Santa Barbara, CA 93106 USA\\
$^{13}$Dipartimento di Fisica, Universit\`a di Milano-Bicocca,\\
Piazza della Scienza 3, I-20126 Milano, Italy\\
$^{14}$Max-Planck-Institut f\"ur Mathematik,
 Vivatsgasse 7,
53111 Bonn, Germany\\
   $^{15}$ Department of Mathematics,  Uppsala University, 
   Box 480, SE-75106 Uppsala, Sweden\\
   $^{16}$ C.~N.~Yang Institute for Theoretical Physics, Stony
  Brook University, Stony Brook, NY 11794-3840, USA\\
$^{17}$ Department of Physics, Technion, Haifa, 32000, Israel\\
$^{18}$Joseph Henry Laboratories, Princeton University, Princeton, NJ 08544, USA\\
$^{19}$Kavli Institute for the Physics and Mathematics of the Universe, \\
 University of Tokyo,  Kashiwa, Chiba 277-8583, Japan\\
$^{20}$KITP/UC Santa Barbara\\
$^{21}$Nordita, KTH Royal Institute of Technology and Stockholm University,
Roslagstullsbacken 23, SE-106 91 Stockholm, Sweden }

\def\abstractvalue{This is the foreword to the special volume
on localization techniques in quantum field theory.
  The summary of individual chapters is given and their interrelation is discussed. 
}

\def\preprintvalue{}

\ifx\ifLONG\undefined
\documentclass[12pt,psamsfonts,reqno]{amsart} 
\newcommand{\chapterauthor}[1]{
\begin{center}
{\bf \normalsize  #1}
\end{center}
}

\newcommand{\chapteraddress}[1]{
\begin{center}
{ \small \it \addressvalue}
\end{center}
}

\newcommand{\chapterabstract}[1]{
\vspace{\baselineskip}
\begin{center}
\textbf{\small Abstract}
\end{center}
#1}

\newcommand{\ifvolume}[2]{\ifx\ifLONG\undefined#2\else#1\fi}

\newcommand{\documentfinish}{
\ifx\ifLONG\undefined
\bibliographystyle{bibreview} 
\bibliography{\IDvalue,review}  
\end{document}
\else
\addcontentsline{toc}{section}{References}
\providecommand{\href}[2]{#2}\begingroup\raggedright\begin{thebibliography}{10}

\bibitem{ContributionPZ}
V.~Pestun and M.~Zabzine, ``Introduction to localization in quantum field
  theory,'' {\em Journal of Physics A} {\bf xx} (2016)  000,
  \href{http://arxiv.org/abs/1608.02953}{{\tt 1608.02953}}.

\bibitem{ContributionPE}
V.~Pestun, ``Review of localization in geometry,'' {\em Journal of Physics A}
  {\bf xx} (2016)  000, \href{http://arxiv.org/abs/1608.02954}{{\tt
  1608.02954}}.

\bibitem{ContributionBL}
F.~Benini and B.~{Le Floch}, ``Supersymmetric localization in two dimensions,''
  {\em Journal of Physics A} {\bf xx} (2016)  000,
  \href{http://arxiv.org/abs/1608.02955}{{\tt 1608.02955}}.

\bibitem{ContributionMO}
D.~Morrison, ``Gromov-Witten invariants and localization,'' {\em Journal of
  Physics A} (2016)  , \href{http://arxiv.org/abs/1608.02956}{{\tt
  1608.02956}}.

\bibitem{ContributionDU}
T.~Dumitrescu, ``An Introduction to Supersymmetric Field Theories in Curved
  Space,'' {\em Journal of Physics A} {\bf xx} (2016)  000,
  \href{http://arxiv.org/abs/1608.02957}{{\tt 1608.02957}}.

\bibitem{ContributionWI}
B.~Willett, ``Localization on three-dimensional manifolds,'' {\em Journal of
  Physics A} {\bf xx} (2016)  000, \href{http://arxiv.org/abs/1608.02958}{{\tt
  1608.02958}}.

\bibitem{ContributionMA}
M.~Mari{\~n}o, ``Localization at large $N$ in Chern-Simons-matter theories,''
  {\em Journal of Physics A} {\bf xx} (2016)  000,
  \href{http://arxiv.org/abs/1608.02959}{{\tt 1608.02959}}.

\bibitem{ContributionPU}
S.~Pufu, ``The F-Theorem and F-Maximization,'' {\em Journal of Physics A} {\bf
  xx} (2016)  000, \href{http://arxiv.org/abs/1608.02960}{{\tt 1608.02960}}.

\bibitem{ContributionDI}
T.~Dimofte, ``Perturbative and nonperturbative aspects of complex Chern-Simons
  Theory,'' {\em Journal of Physics A} {\bf xx} (2016)  000,
  \href{http://arxiv.org/abs/1608.02961}{{\tt 1608.02961}}.

\bibitem{ContributionHO}
K.~Hosomichi, ``$\mathcal{N}=2$ SUSY gauge theories on $S^4$,'' {\em Journal of
  Physics A} {\bf xx} (2016)  000, \href{http://arxiv.org/abs/1608.02962}{{\tt
  1608.02962}}.

\bibitem{ContributionZA}
K.~Zarembo, ``Localization and AdS/CFT Correspondence,'' {\em Journal of
  Physics A} {\bf xx} (2016)  000, \href{http://arxiv.org/abs/1608.02963}{{\tt
  1608.02963}}.

\bibitem{ContributionTA}
Y.~Tachikawa, ``A brief review of the 2d/4d correspondences,'' {\em Journal of
  Physics A} {\bf xx} (2016)  000, \href{http://arxiv.org/abs/1608.02964}{{\tt
  1608.02964}}.

\bibitem{ContributionRR}
L.~Rastelli and S.~Razamat, ``The supersymmetric index in four dimensions,''
  {\em Journal of Physics A} {\bf xx} (2016)  000,
  \href{http://arxiv.org/abs/1608.02965}{{\tt 1608.02965}}.

\bibitem{ContributionQZ}
J.~Qiu and M.~Zabzine, ``Review of localization for 5D supersymmetric gauge
  theories,'' {\em Journal of Physics A} {\bf xx} (2016)  000,
  \href{http://arxiv.org/abs/1608.02966}{{\tt 1608.02966}}.

\bibitem{ContributionMI}
J.~Minahan, ``Matrix models for 5D super Yang-Mills,'' {\em Journal of Physics
  A} {\bf xx} (2016)  000, \href{http://arxiv.org/abs/1608.02967}{{\tt
  1608.02967}}.

\bibitem{ContributionPA}
S.~Pasquetti, ``Holomorphic blocks and the 5d AGT correspondence,'' {\em
  Journal of Physics A} {\bf xx} (2016)  000,
  \href{http://arxiv.org/abs/1608.02968}{{\tt 1608.02968}}.

\bibitem{ContributionKL}
S.~Kim and K.~Lee, ``Indices for 6 dimensional superconformal field theories,''
  {\em Journal of Physics A} {\bf xx} (2016)  000,
  \href{http://arxiv.org/abs/1608.02969}{{\tt 1608.02969}}.

\end{thebibliography}\endgroup

\fi
}

\newcommand{\documentfinishBBL}{
\addcontentsline{toc}{section}{References}
\ifx\ifLONG\undefined
\input{\IDvalue.separate.bbl}
\end{document}
\else
\input{\DIRvalue/\IDvalue.volume.bbl}
\fi
}

\ifx\ifLONG\undefined
\def\volcite#1{Contribution \cite{Contribution#1}}
\else 
\def\volcite#1{Chapter \ref{Chapter#1}}
\fi

\usepackage{ytableau}
\usepackage[margin=1in]{geometry}
\usepackage{amsmath}
\usepackage{amssymb}
\usepackage{amsxtra}
\usepackage{amscd}
\usepackage{xcolor}
\usepackage{tikz-cd}
\usetikzlibrary{matrix,arrows,decorations.pathmorphing}
\usepackage[mathscr]{euscript}
\usepackage{amscd}
\usepackage{slashed}
\usepackage{listings}
\usepackage[inline]{showlabels}
\usepackage{graphicx}
\usepackage{todonotes}
\usepackage[pagebackref=false]{hyperref}
\definecolor{darkblue}{RGB}{0,0,192}
\definecolor{darkgreen}{RGB}{0,192,0}
\hypersetup{colorlinks = true,extension = notused,linkcolor = darkblue,anchorcolor = red,citecolor = darkgreen,filecolor = red,pagecolor = red,urlcolor = darkblue}
\usepackage{iftex}
\ifxetex
        \usepackage{fontspec}
\else
\fi

\newcommand{\libs}[1]{\href{file://localhost/Users/pestun/Dropbox/lib/spires/#1}{\nolinkurl{#1}}}
\newcommand{\libm}[1]{\href{file://localhost/Users/pestun/Dropbox/lib/math/#1}{\nolinkurl{#1}}}
\newcommand{\libt}[1]{\href{file://localhost/Users/pestun/Dropbox/lib/talks/#1}{\nolinkurl{#1}}}
\newcommand{\libb}[1]{\href{file://localhost/Users/pestun/Dropbox/lib/books/#1}{\nolinkurl{#1}}}
\newcommand{\libr}[1]{\href{file://localhost/Users/pestun/Dropbox/lib/research/#1}{\nolinkurl{#1}}}
\newcommand{\libn}[1]{\href{file://localhost/Users/pestun/Dropbox/lib/research/_notes/#1}{\nolinkurl{#1}}}
\newcommand{\libp}[1]{\href{file://localhost/Users/pestun/Dropbox/lib/pestun/#1}{\nolinkurl{#1}}}

\numberwithin{equation}{section}

\begin{document}
\begin{flushright} \small
  \preprintvalue
 \end{flushright}

\begin{center}
{\bf \Large \titlevalue}
\end{center}

\chapterauthor{\authorvalue}
\chapteraddress{\addressvalue}
\chapterabstract{\abstractvalue}

\medskip

\tableofcontents
\else 
\fi

\newcommand{\PZLie}{\mathrm{Lie}}

\section{Summary} 

This is the summary of the special volume ``Localization techniques in
quantum field theories'' which contains 17 individual
chapters.\footnote{The preprint version is available at \url{https://arxiv.org/src/1608.02952/anc/LocQFT.pdf} or \url{http://pestun.ihes.fr/pages/LocalizationReview/LocQFT.pdf}} The
focus of the volume is on the localization technique and its
applications in supersymmetric gauge theories. Although the ideas of
equivariant localization in quantum field theory go back 30 years,
here we concentrate on the recent surge in the subject during the last
ten years. This subject develops rapidly  and thus it is impossible to
have a fully satisfactory overview of the field. This volume took about
two and a half years in making, and during this period some important new
results have been obtained, and it was hard to incorporate all of
them. However we think that it is important to provide an overview and
an introduction to this quickly developing subject. This is important
both for the young researchers, who just enter the field and to
established scientists as well. We have tried to do our best to review
the main results during the last ten years.
   
The volume has two types of chapters, some chapters concentrate on the
localization calculation in different dimensions by itself, and other
chapters concentrate on the major applications of the localization
result. Obviously, such separation is sometimes artificial. The chapters
are ordered roughly according to the dimensions of the corresponding
supersymmetric theories. First, we try to review the
localization calculation in given dimension, and then we move to the
discussion of the major applications.
 
 The volume covers the localization calculations for the supersymmetric theories in dimensions 2,3,4 and 5. The volume 
  discusses the applications of these calculations for theories living up to dimension 6, and  for string/M theories.  
  We have to apologize in advance for omitting 
  from the review the new and important calculations which have appeared 
  during last couple of years. 

This volume is intended to be a single volume where the different
chapters cover the different but related topics within a certain focused
scope.  Some chapters depend on results presented in a different 
 chapter, but the dependency is not a simple linear order. 

\medskip

The whole volume, when published,  could be cited as 
\begin{quote}
V.~Pestun and M. Zabzine, eds.,\\
``Localization techniques in quantum field theories'', \\
{\em Journal of Physics A}  (2016)
\end{quote}
The arXiv preprint version can be accessed from arXiv summary entry
which lists  all authors and links to all 17 individual
contributions, the corresponding citation would be 
\begin{quote}
 \href{http://arxiv.org/abs/1608.02952}{{\tt arXiv:1608.02952}}
\end{quote}
An  individual contribution can be cited by its chapter
number, for exampe    
\begin{quote}  
S.~Pufu, ``The F-Theorem and F-Maximization,''\\
Chapter 8 in  V.~Pestun and M. Zabzine, eds.,\\
``Localization techniques in quantum field  theories'',  \\
{\em Journal of Physics A} (2016) 
\end{quote}
and accessed on arXiv and cited by the arXiv number
\begin{quote}
 \href{http://arxiv.org/abs/1608.02960}{{\tt arXiv:1608.02960}}.
\end{quote}

 \section{Individual chapters}   
 
 Below we summarize the content of each individual contribution/chapter:\\
 
\volcite{PZ}:  ``Introduction to  localization  in quantum field theory''
 (Vasily Pestun and Maxim Zabzine)\\
 
 This is the introductory chapter to the whole volume, outlining its
 scope and reviewing the field as a whole.    
 The chapter discusses shortly the history of equivariant localization both in 
  finite and infinite dimensional setups. The derivation of the finite dimensional Berline-Vergne-Atiyah-Bott formula is given
   in terms of supergeometry. This derivation  is formally generalized to the infinite dimensional setup in the context of supersymmetric 
    gauge theories. The result for supersymmetric theories on
    spheres is presented in a uniform fashion over different
    dimensions, and the related 
    index theorem calculations are reviewed.  The applications of localization techniques are listed and briefly discussed.    \\

\volcite{PE}:  ``Review of localization in geometry'' (Vasily Pestun)\\

This chapter is a short summary of the mathematical aspects of the Berline-Vergne-Atiyah-Bott formula and  Atiyah-Singer index theory. These tools are routinely used throughout the volume. The chapter reviews the definition of equivariant cohomology, 
and its Weyl and Cartan models. The standard characteristic classes and their equivariant versions
    are reviewed.
   The equivariant integration is discussed and the mathematical derivation of  the Berline-Vergne-Atiyah-Bott formula is explained. 
    The Atiyah-Singer index theorems and their equivariant versions are briefly reviewed. 
 \\  
 
 \volcite{BL}:  ``Supersymmetric localization in two dimensions'' (Francesco Benini
 and Bruno Le Floch)\\
 
 This chapter concentrates on the localization techniques for 2d supersymmetric gauge theories and on the major applications 
  of 2d localization results. 
  The main example is the calculation of the partition function for $\mathcal{N}=(2,2)$ gauge theory on $S^2$.  
   Two different approaches are presented, the Coulomb branch
   localization (when the result is written as an integral over  the Cartan subalgebra) and the Higgs branch localization (when the answer is written as a sum). 
     Briefly $\mathcal{N}=(2,2)$ gauge theories on other curved backgrounds are discussed, and the calculation for the hemisphere is presented. 
 The important calculation of the partition function for $\mathcal{N}=(2,2)$ and $\mathcal{N}=(2,0)$ theories on the torus is presented,
  this quantity is known as the elliptic genus. The result is written in
  terms of the Jeffrey-Kirwan residue, which is a higher dimensional 
   analog of the residue operation.  The mathematical aspects of the
   Jeffrey-Kirwan residue operation are briefly explained. 
    As the main application of the localization calculation in 2d, some dualities are discussed; in particular mirror symmetry and 
      Seiberg duality.  
   \\
 
\volcite{MO}: ``Gromov-Witten invariants and localization'' (David Morrison)\\
 
 This chapter concentrates on an important application of 2d localization calculation, see \volcite{BL}. 
  The chapter provides a pedagogical introduction to
  the relation between the genus 0 Gromov-Witten invariants (counting of holomorphic maps) and the localization of 2d gauged linear 
   sigma models. The relation is based on the conjecture which connects the partition function of $\mathcal{N}=(2,2)$ gauge theories on $S^2$ 
    with the Zamolodchikov metric on the conformal manifold of the theory. This relation allows to deduce the Gromov-Witten invariants
    on the Calabi-Yau manifold from the partition function on $S^2$ of the corresponding linear sigma model. This chapter explains 
     this conjecture and reviews the main step of the calculation. 
    \\

\volcite{DU}:  ``An Introduction to supersymmetric field theories in curved space''
 (Thomas Dumitrescu)\\
 
 This chapter addresses the problem of defining  rigid supersymmetric theories on curved backgrounds. 
  The systematic approach to this problem is based on the
  Festuccia-Seiberg work on organizing the background fields into off-shell supergravity multiplets. 
  The chapter concentrates in details 
   on two major examples, $\mathcal{N}=1$ supersymmetric theories in 4d and $\mathcal{N}=2$ supersymmetric theories in 
   3d. The full classification of supersymmetric theories on curved backgrounds can be given for the theories 
    with four or fewer supersymmetry in four or fewer dimensions. 
 \\

\volcite{WI}: ``Localization on three-dimensional manifolds'' (Brian Willett)\\

This chapter provides an introduction to the localization technique for 3d supersymmetric gauge theories. 
The 3d $\mathcal{N}=2$ supersymmetric theories are introduced and their formulation on  curved space is briefly discussed, this is closely 
  related to \volcite{DU}. The calculation of the partition function on
  $S^3$ is presented in great details with the final result presented  as 
an integral over the Cartan sublagebra of the Lie algebra of the gauge group. 
The calculation on the lens spaces, on $S^2 \times S^1$ and
    different applications of these calculations are also discussed. 
The dualities between different gauge theories are briefly
     discussed. The factorization of the result into holomorphic blocks
     is also considered, and in this context the Higgs 
      branch localization is discussed. 
\\
 
\volcite{MA}: ``Localization at large $N$ in Chern-Simons-matter theories'' (Marcos
 Mari{\~n}o)\\
 
 The result of the localization calculation in 3d is given in terms of matrix integrals, see \volcite{WI}. These matrix integrals are 
  complicated and it is not easy to extract  information from this answer. This chapter is devoted to
   the study of 3d matrix models and extracting  physical information from them. The chapter concentrates 
    on the famous  ABJM model which plays a crucial role in the AdS/CFT correspondence.   The M-theory expansion for 
    the  ABJM model is discussed in details and the relation to topological strings is presented. 
     \\
 
\volcite{PU}:  ``The F-Theorem and F-Maximization'' (Silviu Pufu)\\

The partition function on $S^3$ for $\mathcal{N}=2$ supersymmetric gauge theories is written as matrix integrals which depend
 on the different parameters of the theory, see \volcite{WI}. This
 chapter studies the properties of the free energy (minus the logarithm of 
  the sphere partition function),  which is regarded as the measure of the degrees of freedom in the theory.
 In particular the chapter states and explains the F-theorem and F-maximization principles for 3d theories. 
  The F-theorem is a 3d analogue of the Zamolodchikov's c-theorem in 2d and the a-theorem in 4d.  For 3d theories the F-theorem makes 
   a precise statement about the idea 
   that the number of degrees of freedom decreases  along the RG flow. 
\\
 
\volcite{DI}: ``Perturbative and nonperturbative aspects of complex Chern-Simons Theory'' (Tudor Dimofte)\\

This chapter discusses another important application for the
localization calculation in 3d. The chapter starts by briefly reviewing some basic facts 
    about the complex Chern-Simons theory, the main interest is the Chern-Simons theory for $SL(N, \mathbb{C})$.
    There is a short discussion of the 3d/3d correspondence, which states that the partition function of the 
    complex Chern-Simons theory on $M$ is the same as the partition function of a specific supersymmetric gauge theory 
     (whose field content depends on $M$) on the lens space. 
    The chapters finishes with a discussion of the quantum modularity conjecture. 
\\

\volcite{HO}: ``${\mathcal{N}}=2$ SUSY gauge theories on $\mathbf{S}^4$'' (Kazuo Hosomichi)\\

This chapter gives a detailed exposition of the calculation of the partition function and other supersymmetric observables for 
 $\mathcal{N}=2$ supersymmetric gauge theories on $S^4$, both round and squashed. Using  off-shell supergravity,
 the construction of $\mathcal{N}=2$ supersymmetric  theories on squashed $S^4$ is presented. The localization calculation 
  is performed and the determinants are explicitly evaluated using index
  theorems (review in \volcite{PE}).  The inclusion of   supersymmetric observables (Wilson loops, 't Hooft operators and  surface operators) into the localization 
   calculation on $S^4$ is discussed.
   \\
 
\volcite{ZA}: ``Localization and AdS/CFT Correspondence'' (Konstantin Zarembo)\\

One of the major application of the localization calculation on $S^4$ (see \volcite{HO}) is the application to AdS/CFT.
 This chapter is devoted to the study of the matrix models which appear in the calculation on $S^4$ and its application to 
  the AdS/CFT correspondence. Localization offers a unique laboratory for
  the AdS/CFT correspondence, since we are able to 
   explore the supersymmetric gauge theory in non-perturbative
   domain. Using holography the localization computation 
    can be compared to  string theory and supergravity calculations. 
\\
 
\volcite{TA}:  ``A brief review of the 2d/4d correspondences'' (Yuji Tachikawa)\\

From the perspective of
the $\mathcal{N}=(0,2)$ self-dual 6d theory, this chapter explains the
2d/4d correspondence (AGT), considering the 6d theory on a product of 2d and
4d manifold.   This correspondence relates the 4d computations for
supersymmetric gauge theories of class $\mathcal{S}$, obtained by
compactification of the 6d theory
on the 2d manifold, to 2d computations in 2d theory obtained by
compactification of 6d theory on the 4d manifold.  The chapter  starts by reviewing basic facts about 2d q-deformed Yang-Mills theory and the Liouville theory.
 The main building block of rank 1 theories considered in the chapter is the
   trifundamental multiplet coupled with $SU(2)$ gauge fields. 
      The partition function on $S^1 \times S^3$  for such a 4d theory
      is computed in 2d  by q-deformed Yang-Mills and the partition
      function on $S^4$ is computed in 2d by the Liouville theory. 
  \\
 
\volcite{RR}: ``The supersymmetric index in four dimensions'' (Leonardo Rastelli and
 Shlomo Razamat) \\
 
 This chapter studies the partition function on $S^3 \times S^1$ for ${\mathcal N}=1$ superymmetric theories in 4d, also known as the 4d supersymmetric index. The chapter starts by defining the supersymmetric index and
reviewing combinatorial tools to compute it in theories with a Lagrangian description. After illustrating some basic properties of the index in the simple setting of supersymmetric sigma models,
the chapter turns to the discussion  of the index of supersymmetric
gauge theories, emphasizing physical applications. The index contains
useful information about the spectrum of shortened multiplets, and how to extract this information is discussed in some detail. The most important application of the index, as a powerful tool for checking non-perturbative dualities between supersymmetric gauge theories, is illustrated in several examples. The  last part of the chapter considers several interesting limits of the supersymmetric index.
  \\
 
\volcite{QZ}: ``Review of localization for 5d supersymmetric gauge theories'' (Jian
Qiu and Maxim Zabzine)\\

The chapter provides the introduction to localization calculation for $\mathcal{N}=1$ supersymmetric gauge theories 
 on  toric Sasaki-Einstein manifolds, for example on a five-sphere $S^5$. 
  The chapter starts by recalling basic facts about supersymmetry and supersymmetric gauge theories in 
 flat 5d space. Then the construction of the supersymmetric gauge theory on the Sasaki-Einstein manifolds
    is explicitly given.  Using the field redefinition, the supersymmetry transformations are rewritten in terms 
     of differential forms, thus making geometrical aspects of the localization more transparent. 
      For  toric Sasaki-Einstein manifolds the localization calculation
      can be carried out completely, the calculation 
       of determinants is given and the full partition function is conjectured. The chapter ends with comments 
        about deducing the flat space results from the curved result. 
    \\
 
\volcite{MI}: ``Matrix models for 5d super Yang-Mills'' (Joseph Minahan)\\
 
 The result of 5d localization calculation is given in terms of complicated matrix models, see \volcite{QZ}.
  This chapter studies the resulting matrix models. The basic properties of 
   the matrix models are described and the 't Hooft limit is analyzed
   for $\mathcal{N}=1^{*}$ theory (a vector multiplet plus 
   a hypermultiplet in the adjoint representation). For large 't Hooft coupling the free energy 
    behaves as $N^3$ for $U(N)$ gauge theory and the corresponding supergravity analysis is performed.
     This analysis support  the idea that the non-perturbative
     completion of 5d theory is the 6d $\mathcal{N}=(2,0)$ 
      superconformal field theory. 
      \\
 
\volcite{PA}:  ``Holomorphic blocks and the 5d AGT correspondence'' (Sara Pasquetti)\\

This chapter further studies partition functions from 2d to 5d. In particular it concentrates on 
 the idea that the partition function on a compact manifold can be built up from basic blocks, so-called 
  holomorphic blocks. The main point is that the geometric decomposition of the compact manifold should 
   have its counterpart in the appropriate decomposition of the partition function.  These factorization properties 
    are reviewed in different dimensions. The rest of the chapter
    concentrates on a 5d version of the  AGT correspondence.  
\\

\volcite{KL}: ``Indices for 6 dimensional superconformal  field theories'' (Seok Kim and
Kimyeong Lee)\\

This chapter deals with the 6d $(2,0)$ superconformal field theory. This theory cannot be accessed directly, 
 but it is related to many other supersymmetric gauge theories, e.g. it is believed to be the UV-completion of
  maximally supersymmetric 5d gauge theory.  The relation between 5d partition function and 6d supersymmetric 
   index is discussed in details in this chapter.  
\\

\newpage
\section{Volume structure}

The different chapters are related to each other and the relation is not
a simple linear relation, which can be shown by their ordering in the volume. Below we provide the graphical relation between different chapters.
  This diagram\footnote{Special thanks to Yuji Tachikawa for the final design of the diagram} gives the general idea. \\

\includegraphics[scale=0.4]{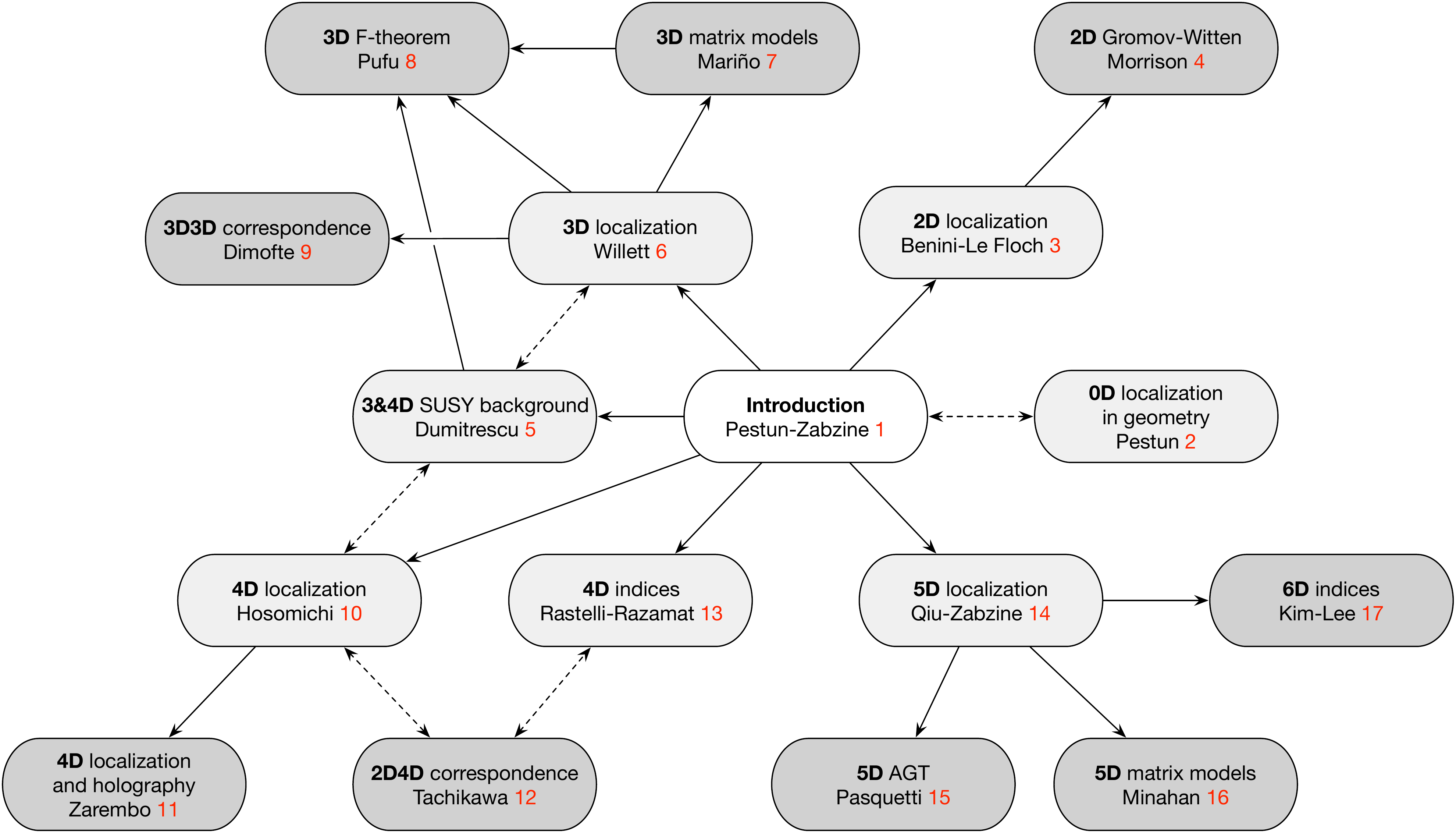}


\documentfinish
